\documentclass{article}
\usepackage{amssymb}
\usepackage{cite}
\usepackage{graphicx}
\usepackage{dcolumn}

\begin{document}

\date{}
\title{An ubiquitous three-term recurrence relation}
\author{Paolo Amore\thanks{%
e--mail: paolo@ucol.mx} \\
Facultad de Ciencias, CUICBAS, Universidad de Colima,\\
Bernal D\'{\i}az del Castillo 340, Colima, Colima,Mexico \\
and \\
Francisco M. Fern\'andez\thanks{%
e--mail: fernande@quimica.unlp.edu.ar} \\
INIFTA, Divisi\'{o}n Qu\'{\i}mica Te\'{o}rica,\\
Blvd. 113 y 64 (S/N), Sucursal 4, Casilla de Correo 16,\\
1900 La Plata, Argentina}
\maketitle

\begin{abstract}
We solve an eigenvalue equation that appears in several papers about a wide
range of physical problems. The Frobenius method leads to a three-term
recurrence relation for the coefficients of the power series that, under
suitable truncation, yields exact analytical eigenvalues and eigenfunctions
for particular values of a model parameter. From these solutions some
researchers have derived a variety of predictions like allowed angular
frequencies, allowed field intensities and the like. We also solve the
eigenvalue equation numerically by means of the variational Rayleigh-Ritz
method and compare the resulting eigenvalues with those provided by the
truncation condition. In this way we prove that those physical predictions
are merely artifacts of the truncation condition.
\end{abstract}

\section{Introduction}

$\label{sec:intro}$

In a series of papers, several authors discuss a wide variety of physical
models in cylindrical coordinates that, after some suitable transformations
of the main dynamical (or eigenvalue) equation, can be reduced to an
eigenvalue equation in the radial variable with Coulomb (or Coulomb--like)
and harmonic (or harmonic-like) interactions. Through further application of
the Frobenius (power-series) method they obtain a three-term recurrence
relation. They state that in order to have finite solutions (or normalizable
ones) the power series should terminate and they place a suitable truncation
condition for that purpose. As a result of this truncation those authors
invariably draw the conclusion that some model parameters, like the
intensity of a magnetic field or the oscillator frequency, for example,
should be discrete. In other words, they claim that the eigenvalue equation
has square-integrable solutions for some particular values (allowed values)
of such parameters. For example, Ver\c{c}in\cite{V91} derives an exact
solution to the problem of two identical charged anyons moving in a plane
under the influence of a static uniform magnetic field perpendicular to that
plane. He argues that there are bound states if and only if the series
terminates, which occurs only for certain discrete values of the magnetic
field. Later, Myrheim et al\cite{MHV92} discussed Ver\c{c}in's results with
more detail. Furtado et al\cite{FDMBB94} discuss the influence of a
disclination on the spectrum of an electron or a hole in a magnetic field in
the framework of the theory of defects and three-dimensional gravity of
Katanaev and Volovich\cite{KV92}. In this case the cyclotron frequency and
the magnetic field should depend on the quantum numbers. Bakke and Moraes%
\cite{BM12} introduce a geometric model to explain the origin of the
observed shallow levels in semiconductors threaded by a dislocation density
and find allowed values of the oscillator frequency or of the constant $k$
associated to the momentum along the $z$-axis. Bakke and Beilish\cite{BB12}
obtain the bound states for a non-relativistic spin-half neutral particle
under the influence of a Coulomb-like potential induced by the Lorentz
symmetry breaking effects. They claim to present a new possible scenario of
studying the Lorentz symmetry breaking effects on a non-relativistic quantum
system defined by a fixed space-like vector field parallel to the radial
direction interacting with a uniform magnetic field along the $z$-axis. They
also discuss the influence of a Coulomb-like potential induced by Lorentz
symmetry violation effects on a two-dimensional harmonic oscillator and find
allowed values of the cyclotron frequency. Bakke\cite{B14} discusses a model
that consists of the interaction between a moving electric quadrupole moment
and a magnetic field and also adds a two-dimensional harmonic-oscillator
potential thus obtaining allowed values of the oscillator frequency. Bakke%
\cite{B14b} studies the bound states of a quantum-mechanical model given by
the interaction between the electric quadrupole of a moving particle and an
electric field. In two other models the author adds a harmonic potential and
a linear plus harmonic potential and also finds allowed oscillator
frequencies. Bakke and Belich\cite{BB14} study the effects of the Lorentz
symmetry violation in the non-relativistic quantum dynamics of a spin-1/2
neutral particle interacting with external fields confined to a
two-dimensional quantum ring and also finds oscillator frequencies that
depend on the quantum numbers. Fonseca and Bakke\cite{FB15} propose a model
for the interaction of a magnetic quadrupole moment with electric and
magnetic fields and in a second model they add a harmonic-oscillator
potential finding that the angular frequency depends on the quantum numbers.
Bakke and Furtado\cite{BF15} study the influence of a Coulomb-type potential
on the Klein-Gordon oscillator and find that the angular frequency should
depend on the quantum numbers.

The purpose of this paper is to study the radial eigenvalue equation derived
in those papers and investigate to which extent the truncation of the power
series by means of the tree-term recurrence relation affects the physical
conclusions drawn by their authors. In section~\ref{sec:time-dependent} we
outline the main equation solved in the papers mentioned above. In section~%
\ref{sec:three-term-rec} we solve the radial eigenvalue equation by means of
the Frobenius method and truncation through a three-term recurrence relation
for the series coefficients in order to derive analytical solutions. By
means of a reliable variational method we also obtain accurate numerical
eigenvalues that are compared with the analytical ones. Finally, in section~%
\ref{sec:conclusions} we summarize the main results and draw conclusions.

\section{The time-dependent equation}

\label{sec:time-dependent}

In several of the papers discussed here the starting point is a
time-dependent quantum-mechanical equation of the form\cite%
{BM12,BB12,B14,B14b,BB14,FB15}%
\begin{equation}
i\frac{\partial \psi }{\partial t}=H\psi ,  \label{eq:time_dep_general}
\end{equation}%
where the Hamiltonian operator $H=H\left( \rho ,\partial _{\rho },\partial
_{\varphi },\partial _{z}\right) $, $\partial _{q}=\partial /\partial q$,
is given in cylindrical coordinates $0<\rho <\infty $, $0\leq \varphi \leq
2\pi $, $-\infty <z<\infty $. Upon choosing the particular solution%
\begin{equation}
\psi (t,\rho ,\varphi ,z)=e^{-i\mathcal{E}t}e^{ij\varphi }e^{ikz}R(\rho ),
\label{eq:particular_sol_general}
\end{equation}%
where $j=l=0\pm 1\pm 2,\ldots $ in some cases\cite{B14,B14b,FB15,BF15}, $%
j=l+1/2$ in others\cite{BM12,BB12,BB14} and $-\infty <k<\infty $ the authors
derive a eigenvalue equation for $R(\rho )$:%
\begin{equation}
H_{jk}\left( \rho ,\partial _{\rho }\right) R(\rho )=\mathcal{E}R(\rho ).
\label{eq:eigen_R_gen}
\end{equation}

In most of those papers the authors state that they are looking for bound
states although it is plain that their models do not support such kind of
solutions. As a matter of fact, a bound state requires that%
\begin{equation}
\int \int \int \left\vert \psi (t,\rho ,\varphi ,z)\right\vert ^{2}\rho
\,d\rho \,d\varphi \,dz<\infty ,  \label{eq:bound-state_def_psi}
\end{equation}%
but in the general example outlined above the improper integral over $z$ is
divergent. Besides, the energy depends on $-\infty <k<\infty $ which clearly
shows that the spectrum is continuous. This is one of the conceptual errors
in those papers. Other authors consider a motion in a plane where there are
truly bound states\cite{V91,MHV92,BF15}. In what follows we assume that the
motion of the particle in the three-dimensional space can be restricted to a
fictitious motion in the $x-y$ plane so that we can truly speak of bound
states.

As an illustrative example we consider the influence of a disclination on
the spectrum of an electron or a hole in a magnetic field in the framework
of the theory of defects and three-dimensional gravity of Katanaev and
Volovich\cite{KV92} discussed by Furtado et al\cite{FDMBB94}. The model
Hamiltonian derived from the metric%
\begin{equation}
ds^{2}=dz^{2}+d\rho ^{2}+\alpha ^{2}\rho ^{2}d\phi ^{2},
\end{equation}%
in cylindrical coordinates and the interaction between the charge of the
particle $q$ and the magnetic field $\mathbf{B}$ is%
\begin{equation}
H=-\frac{\hbar ^{2}}{2m^{\ast }}\left( \frac{1}{\rho }\frac{\partial }{%
\partial \rho }\rho \frac{\partial }{\partial \rho }+\frac{1}{\alpha
^{2}\rho ^{2}}\frac{\partial ^{2}}{\partial \phi ^{2}}+\frac{\partial ^{2}}{%
\partial z^{2}}\right) +\frac{i\hbar qB}{2\alpha ^{2}m^{\ast }c}\frac{%
\partial }{\partial \phi }+\frac{q^{2}B^{2}}{8m^{\ast }c^{2}\alpha ^{2}}\rho
^{2},  \label{eq:H_FDMBB94}
\end{equation}%
where the precise meaning of each parameter is given in the authors' paper%
\cite{FDMBB94}. The authors also add the self-interaction term%
\begin{equation}
\bar{U}=\frac{q^{2}}{4\pi \epsilon }\frac{\kappa (p)}{\rho },
\end{equation}%
to the Hamiltonian operator in the Schr\"{o}dinger equation $H\psi =E\psi .$

We can easily derive a dimensionless equation by means of a well-known
systematic procedure\cite{F20}. We define the dimensionless coordinates%
\begin{equation}
\tilde{\rho}=\frac{\rho }{L},\,\tilde{z}=\frac{z}{L},\,L=\sqrt{\frac{%
2c\alpha \hbar }{|q|B}},
\end{equation}%
in terms of the unit of length $L$ and the dimensionless Hamiltonian operator%
\begin{eqnarray}
\tilde{H} &=&\frac{2m^{\ast }L^{2}}{\hbar ^{2}}H=-\left( \frac{1}{\tilde{\rho%
}}\frac{\partial }{\partial \tilde{\rho}}\tilde{\rho}\frac{\partial }{%
\partial \tilde{\rho}}+\frac{1}{\alpha ^{2}\tilde{\rho}^{2}}\frac{\partial
^{2}}{\partial \phi ^{2}}+\frac{\partial ^{2}}{\partial \tilde{z}^{2}}%
\right) +\frac{2iq}{\alpha |q|}\frac{\partial }{\partial \phi }+\tilde{\rho}%
^{2}+\frac{a}{\tilde{\rho}},  \nonumber \\
a &=&\frac{m^{\ast }q^{2}\kappa }{\pi \epsilon \hbar ^{2}}\sqrt{\frac{%
c\alpha \hbar }{2|q|B}},  \label{eq:H_dim_FDMBB94}
\end{eqnarray}%
where $\hbar ^{2}/\left( 2m^{\ast }L^{2}\right) $ is a natural unit of
energy. If we propose the particular solution%
\begin{equation}
\psi (\tilde{\rho},\varphi ,\tilde{z})=e^{il\varphi }e^{ik\tilde{z}}R(\tilde{%
\rho}),
\end{equation}%
the resulting differential equation for $R(\tilde{\rho})$ is%
\begin{eqnarray}
&&-\left( \frac{1}{\tilde{\rho}}\frac{\partial }{\partial \tilde{\rho}}%
\tilde{\rho}\frac{\partial }{\partial \tilde{\rho}}-\frac{l^{2}}{\alpha ^{2}%
\tilde{\rho}^{2}}\right) R+\tilde{\rho}^{2}R+\frac{a}{\tilde{\rho}}R=WR,
\nonumber \\
&&W=\frac{4mc\alpha E}{\hslash |q|B}-k^{2}+\frac{2ql}{\alpha |q|}.
\label{eq:eigen_eq_R_FDMBB94}
\end{eqnarray}%
Notice that present parameter $a$ is exactly the parameter $b$ in Furtado et
al's paper\cite{FDMBB94}. In other articles the authors simply state that
they resort to units such that $\hbar =c=1$\cite%
{BM12,BB12,B14,B14b,BB14,FB15,BF15} but this non-rigorous way of choosing
suitable units was recently criticized in a pedagogical paper\cite{F20}.

\section{The three-term recurrence relation}

\label{sec:three-term-rec}

By means of suitable transformations of the eigenvalue equation (\ref%
{eq:eigen_R_gen}) the authors mentioned above derive an eigenvalue equation
of the form\cite{V91,MHV92,FDMBB94,BM12,BB12,B14,B14b,BB14,FB15,BF15}%
\begin{eqnarray}
\hat{L}R &=&WR,  \nonumber \\
\hat{L} &\equiv &-\frac{d^{2}}{d\xi ^{2}}-\frac{1}{\xi }\frac{d}{d\xi }+%
\frac{\gamma ^{2}}{\xi ^{2}}-\frac{a}{\xi }+\xi ^{2},  \label{eq:eigen_eq_R}
\end{eqnarray}%
where $\gamma $ and $a$ are real constants and, in general, $\gamma $
depends on the rotational quantum number $l$. A particular example was
derived in the preceding section.

By means of the ansatz
\begin{equation}
R(\xi )=\xi ^{|\gamma |}e^{-\frac{\xi ^{2}}{2}}P(\xi ),\,P(\xi
)=\sum_{j=0}^{\infty }c_{j}\xi ^{j},  \label{eq:R_series}
\end{equation}%
we obtain a three-term recurrence relation for the coefficients $c_{j}$:%
\begin{eqnarray}
c_{j+2} &=&-\frac{a}{(j+2)(j+2|\gamma |+2)}c_{j+1}+\frac{2j+2|\gamma |+2-W}{%
(j+2)(j+2|\gamma |+2)}c_{j},  \nonumber \\
j &=&-1,0,1,\ldots ,\;c_{-1}=0,\;c_{0}=1.  \label{eq:rec_rel_gen}
\end{eqnarray}%
In the papers just mentioned the authors state, in one way or another, that
in order to obtain bound states one has to force the termination conditions%
\begin{equation}
W=2n+2|\gamma |+2,\,c_{n+1}=0,\,n=1,2,\ldots .  \label{eq:trunc_cond}
\end{equation}%
Clearly, under such conditions $c_{j}=0$ for all $j>n$ and $P(\xi )$ reduces
to a polynomial of degree $n$. In this way, they obtain analytical
expressions for the eigenvalues $W_{n,l}=2n+2|\gamma |+2$ and the radial
eigenfunctions $R_{n,l}(\xi )$\cite%
{V91,MHV92,FDMBB94,BM12,BB12,B14,B14b,BB14,FB15,BF15}. For the sake of
clarity and generality, in this section we use $\gamma $ instead of $l$ as
an effective quantum number because the form of $\gamma $ is not the same in
all those papers. Besides, we will also include the truncation condition for
$n=0$ although in this case the only solution is $a=0$ and the problem
reduces to the exactly solvable harmonic oscillator.

For example, when $n=1$ the truncation condition (\ref{eq:trunc_cond}) yields%
\begin{equation}
W_{1,\gamma }=2\left( |\gamma |+2\right) ,\,a_{1,\gamma }^{\left( 1\right)
}=-\sqrt{2\left( 2|\gamma |+1\right) },\,a_{1,\gamma }^{\left( 2\right) }=%
\sqrt{2\left( 2|\gamma |+1\right) }.  \label{eq:W_a_n=1}
\end{equation}%
There is no doubt that we have obtained only one eigenvalue $W=W_{1,\gamma }$
for two particular values of $a=a_{1,\gamma }^{(i)}$, $i=1,2$, that depend
on the chosen value of $\gamma $. In the general case, we obtain the same
value $W_{n,\gamma }$ for each of the particular roots $a_{n,\gamma }^{(k)}$%
, $k=1,2,\ldots ,n+1$, $a_{n,\gamma }^{(k)}<a_{n,\gamma }^{(k+1)}$, of $%
c_{n+1}=0$ (including $a=0$ as a possible solution). For example, we have
three real roots for $n=2$
\begin{eqnarray}
W_{2,\gamma } &=&2\gamma +6,\,  \nonumber \\
a_{2,\gamma }^{\left( 1\right) } &=&-2\sqrt{4\gamma +3},\,a_{2,\gamma
}^{\left( 2\right) }=0,\,a_{2,\gamma }^{\left( 3\right) }=2\sqrt{4\gamma +3},
\label{eq:W_a_n=2}
\end{eqnarray}%
and four for $n=3$
\begin{eqnarray}
W_{3,\gamma } &=&2\gamma +8,\,  \nonumber \\
a_{2,\gamma }^{\left( 1\right) } &=&-\sqrt{2}\sqrt{10\left( \gamma +1\right)
+\sqrt{64\gamma ^{2}+128\gamma +73}},\,  \nonumber \\
a_{2,\gamma }^{\left( 2\right) } &=&-\sqrt{2}\sqrt{10\left( \gamma +1\right)
-\sqrt{64\gamma ^{2}+128\gamma +73}}  \nonumber \\
a_{2,\gamma }^{\left( 3\right) } &=&\sqrt{2}\sqrt{10\left( \gamma +1\right) -%
\sqrt{64\gamma ^{2}+128\gamma +73}},\,  \nonumber \\
a_{2,\gamma }^{\left( 4\right) } &=&\sqrt{2}\sqrt{10\left( \gamma +1\right) +%
\sqrt{64\gamma ^{2}+128\gamma +73}}.  \label{eq:W_a_n=3}
\end{eqnarray}

It is obvious to anybody familiar with conditionally solvable
quantum-mechanical models\cite{D88, BCD17} (and references therein) that the
approach just described does not produce all the eigenvalues of the operator
$\hat{L}$ for a given set of values of $\gamma $ and $a$ but only those
states with polynomial solutions for $P(\xi )$. These particular eigenvalues
$W_{n,\gamma }=2n+2|\gamma |+2$, $n=0,1,\ldots $ are related to the harmonic
oscillator ones and each of them corresponds to a set of particular values
of $a$, namely $a_{n,\gamma }^{(k)}$. On the other hand, if we solve the
eigenvalue equation (\ref{eq:eigen_eq_R}) in a proper way we obtain an
infinite set of eigenvalues $W_{\nu ,\gamma }(a)$, $\nu =0,1,2,\ldots $ for
each set of real values of $a$ and $\gamma $. The condition that determines
these allowed values of $W$ is that the corresponding radial eigenfunctions $%
R(\xi )$ are square integrable
\begin{equation}
\int_{0}^{\infty }\left\vert R(\xi )\right\vert ^{2}\xi \,d\xi <\infty ,
\label{eq:bound-state_def_xi}
\end{equation}%
as shown in any textbook on quantum mechanics\cite{CDL77}. Notice that $\nu $
is the actual radial quantum number (that labels the eigenvalues in
increasing order of magnitude) whereas $n$ is just a positive integer that
labels some particular solutions with polynomial factors $P(\xi )$. In other
words: $n$ is a fictitious quantum number used in those earlier papers\cite%
{V91,MHV92,FDMBB94,BM12,BB12,B14,B14b,BB14,FB15,BF15}.

The true eigenvalues $W_{\nu ,\gamma }(a)$ of equation (\ref{eq:eigen_eq_R})
are decreasing functions of $a$ as follows from the Hellmann-Feynman theorem%
\cite{CDL77,P68}%
\begin{equation}
\frac{\partial W}{\partial a}=-\left\langle \frac{1}{\xi }\right\rangle <0.
\label{eq:HFT}
\end{equation}%
From this expression we can draw a simple conclusion: for a given value of $n
$ the pair $\left( a_{n,\gamma }^{(1)},W_{n,\gamma }\right) $ is a point on
the true ground-state curve, $W_{0,\gamma }(a)$, $\left( a_{n,\gamma
}^{(2)},W_{n,\gamma }\right) $ is a point on the true first-excited-state
curve $W_{1,\gamma }(a)$, and so on.

The reflection symmetry of $W$ with respect to $a$ suggested by the
analytical energy eigenvalues shown in those earlier papers\cite%
{V91,MHV92,FDMBB94,BM12,BB12,B14,B14b,BB14,FB15,BF15} is fictitious, the
analysis above clearly shows, for example, that: $W_{0,0}\left( -\sqrt{2}%
\right) =W_{1,0}\left( \sqrt{2}\right) =4$, $W_{0,1}\left( -\sqrt{6}\right)
=W_{1,1}\left( \sqrt{6}\right) =6$, $W_{0,0}\left( -\sqrt{12}\right)
=W_{2,0}\left( \sqrt{12}\right) =6$ (using the correct quantum number $\nu $%
). The fact that $W_{\nu ,\gamma }\left( -|a_{n,\gamma }|\right) =W_{\nu
^{\prime },\gamma }\left( |a_{n,\gamma }|\right) $ for $\nu <\nu ^{\prime }$
is a consequence of the Hellmann-Feynman theorem (\ref{eq:HFT}) just
mentioned.

The eigenvalue equation (\ref{eq:eigen_eq_R}) cannot be solved exactly in
the general case. In order to obtain sufficiently accurate eigenvalues of
the operator $\hat{L}$ we resort to the reliable Rayleigh-Ritz variational
method that is well known to yield increasingly accurate upper bounds to all
the eigenvalues of the Schr\"{o}dinger equation\cite{P68} (and references
therein). For simplicity we choose the basis set of non-orthogonal functions
$\left\{ u_{j}(\xi )=\xi ^{|\gamma |+j}e^{-\frac{\xi ^{2}}{2}%
},\;j=0,1,\ldots \right\} $. We test the accuracy of these results by means
of the powerful Riccati-Pad\'{e} method\cite{FMT89a}.

\begin{figure}[tbp]
\begin{center}
\includegraphics[width=9cm]{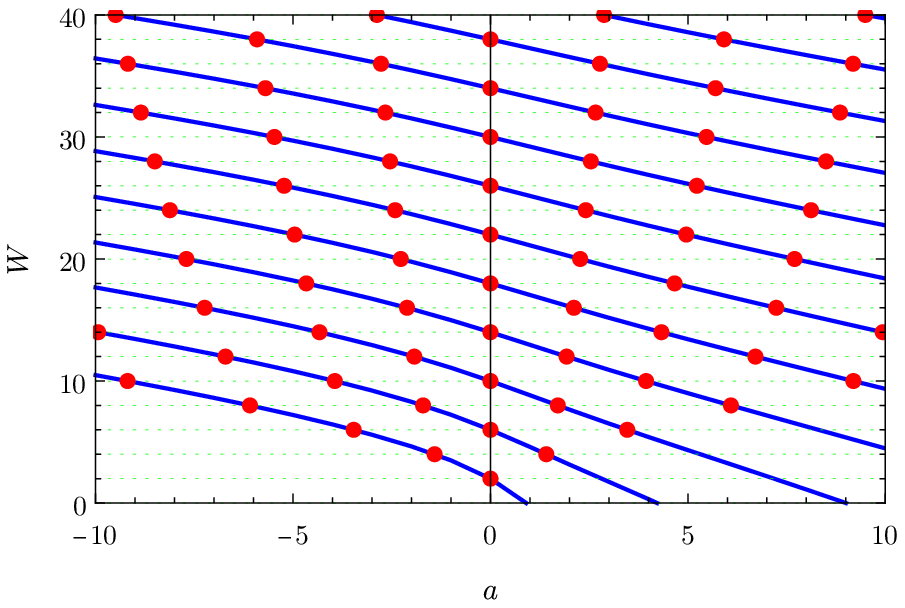}
\end{center}
\caption{Eigenvalues for $\protect\gamma=0$}
\label{fig:W0}
\end{figure}

\begin{figure}[tbp]
\begin{center}
\includegraphics[width=9cm]{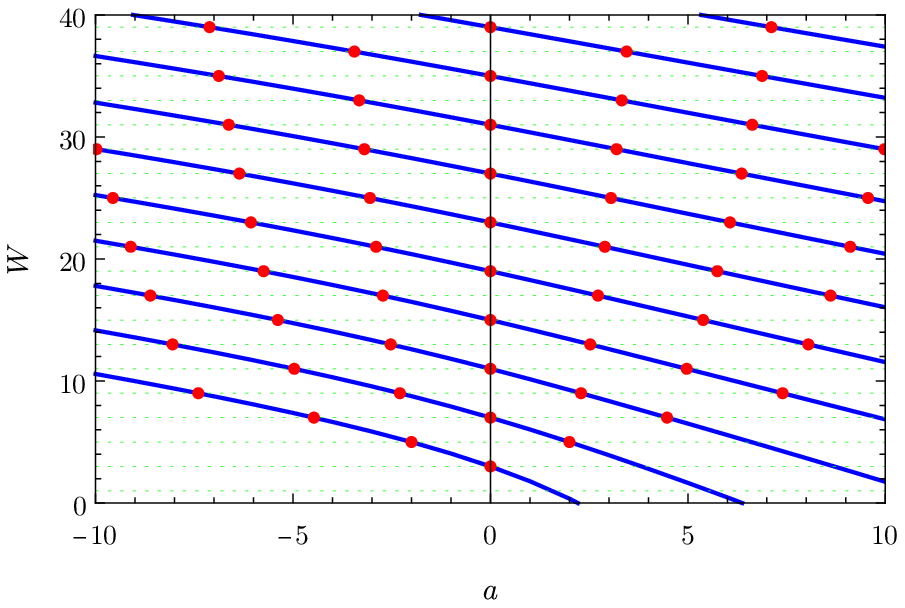}
\end{center}
\caption{Eigenvalues for $\protect\gamma=1/2$}
\label{fig:W05}
\end{figure}

\begin{figure}[tbp]
\begin{center}
\includegraphics[width=9cm]{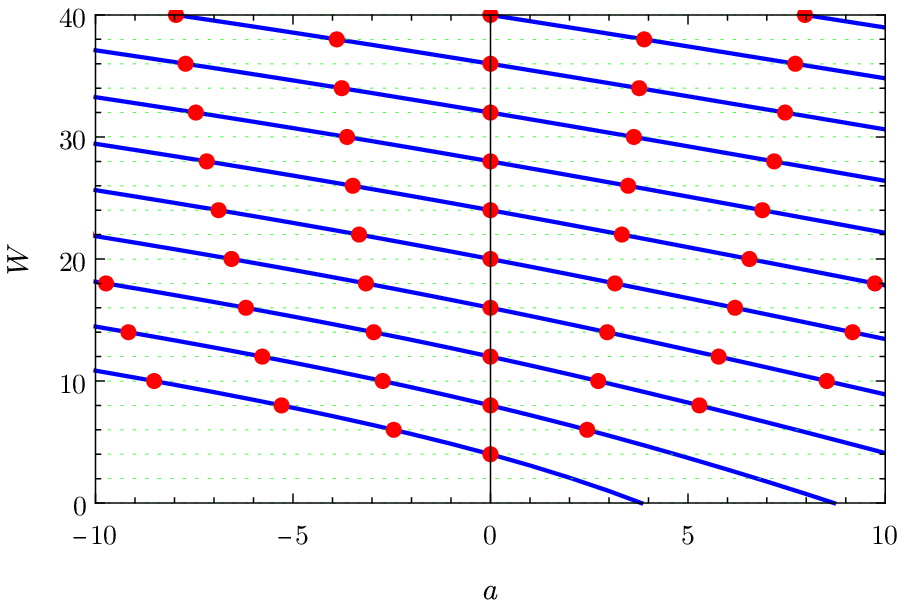}
\end{center}
\caption{Eigenvalues for $\protect\gamma=1$}
\label{fig:W1}
\end{figure}

Figures \ref{fig:W0}, \ref{fig:W05} and \ref{fig:W1} show present numerical
eigenvalues $W_{\nu ,\gamma }$ for $\gamma =0$, $\gamma =1/2$, and $\gamma =1
$, respectively. Every blue, continuous, line is a curve $W_{\nu ,\gamma }(a)
$ and the red circles indicate pairs $\left( a_{n,\gamma }^{(k)},W_{n,\gamma
}\right) $ coming from the artificial truncation condition (\ref%
{eq:trunc_cond}). These figures clearly confirm the theoretical analysis
carried above.

The quantization of the parameter $a$ (allowed values of $a$, or its
dependence on the quantum numbers) is a mere artifact of the truncation
condition (\ref{eq:trunc_cond}) and does not exhibit any physical meaning.
This fact is more than evident in the light of present numerical
calculations. In other words, the existence of allowed values of the field
strength, oscillator frequency, etc, claimed in those earlier papers\cite%
{V91,MHV92,FDMBB94,BM12,BB12,B14,B14b,BB14,FB15,BF15} is merely an artifact
of that unnecessary truncation condition. Consider, for example, the
eigenvalue equation (\ref{eq:eigen_eq_R_FDMBB94}) from which the artificial
quantization of the parameter $a=a_{n,l}$ may be interpreted, according to
equation (\ref{eq:H_dim_FDMBB94}), as the existence of allowed field
strengths $B=B_{n,l}$\cite{FDMBB94}. Present numerical results clearly show
that there are square-integrable solutions for any $B\neq B_{n,l}$. To be
clearer: the set of bound states produced by the truncation condition (\ref%
{eq:trunc_cond}) is contained in the set of all bound states. The majority
of bound states do not satisfy such arbitrary, unnecessary condition.

\section{Conclusions}

\label{sec:conclusions}

Since 1991 several authors have been discussing a wide variety of physical
models that lead to either dynamic or time-independent equations that can be
reduced to a three-term recurrence relation by means of the Frobenius method%
\cite{V91,MHV92,FDMBB94,BM12,BB12,B14,B14b,BB14,FB15,BF15}. By means of an
unnecessary truncation condition they obtain particular solutions (\ref%
{eq:R_series}) with factors $P(\xi )$ that are polynomial functions of the
radial coordinate. These particular solutions take place for particular
values of one or another model parameter ($a$ in the present case). Since
the authors appear to believe that these particular solutions are the only
ones allowed they conjecture that there are allowed angular frequencies or
allowed field intensities and the like. In this paper we show
that all those physical conclusions are mere artifacts of the unnecessary
truncation condition that only produces some particular solutions without
any physical relevance. The only interest in them is purely academic as they
are exact solutions of a problem that is not exactly solvable\cite{D88,
BCD17}. Present numerical calculations, illustrated by means of three
figures, reveal that there are square-integrable solutions for any value of
the dimensionless parameter $a$ (blue lines). Those figures also show the
particular arbitrary eigenvalues (red circles) coming from the truncation
condition  that have been interpreted as if there were allowed angular
frequencies, field intensities, etc. Those particular eigenvalues and their
corresponding model parameters have no relevant role in the physics of the
problem, except that they correspond to polynomial factors $P(\lambda )$.

\end{document}